\documentclass[nojournal]{wiley2sp} 
\usepackage{amsmath}
\usepackage{bm}
\usepackage{mhchem}

\usepackage{siunitx}

\newcommand{\MoSS}{\ce{MoS2}}

\begin{document}

\title{Material transfer and contact optimization in MoS\textsubscript{2}
nanotube devices}

\author{%
  Robin T. K. Schock\textsuperscript{\Ast,\textsf{\bfseries 1}},
  Stefan Obloh\textsuperscript{\textsf{\bfseries 1}},
  Jonathan Neuwald\textsuperscript{\textsf{\bfseries 1}},
  Matthias Kronseder\textsuperscript{\textsf{\bfseries 1}},
  Wolfgang Möckel\textsuperscript{\textsf{\bfseries 1}}, 
  Matja\v{z} Malok\textsuperscript{\textsf{\bfseries 2}},
  Luka Pirker\textsuperscript{\textsf{\bfseries 2,3}},
  Maja Rem\v{s}kar\textsuperscript{\textsf{\bfseries 2}}, 
  Andreas K. Hüttel\textsuperscript{\textsf{\bfseries 1}}
}

\mail{e-mail
  \textsf{andreas.huettel@ur.de}}

\institute{%
  \textsuperscript{1}\,Institute for Experimental and Applied Physics, University of Regensburg, 93053 Regensburg, Germany \\
  \textsuperscript{2}\,Solid State Physics Department, Institute Jo\v{z}ef Stefan, 1000 Ljubljana, Slovenia \\
  \textsuperscript{3}\,J. Heyrovský Institute of Physical Chemistry, v.v.i.,
  Czech Academy of Sciences, 18223 Prague, Czech Republic
}

\keywords{TMDC, MoS2, nanotubes, transfer, contact, optimization}

\abstract{\bf%
While the promise of clean and defect-free MoS\textsubscript{2} nanotubes as
quantum electronic devices is obvious, ranging from strong spin-orbit
interaction to intrinsic superconductivity, device fabrication still poses
considerable challenges. Deterministic transfer of transition metal
dichalcogenide nanomaterials and transparent contacts to the nanomaterials are
nowadays highly active topics of research, both with fundamental research and
applications in mind. Contamination from transport agents as well as surface
adsorbates and surface charges play a critical role for device performance.
Many techniques have been proposed to address these topics for transition metal
dichalcogenides in general. Here, we analyse their usage for the transfer based 
fabrication of \ce{MoS2} nanotube devices. Further, we compare different 
contact materials in order to avoid the formation of a Schottky barrier.
}

\maketitle

\section{Introduction}

Since the first experimental isolation of graphene 
\cite{science-novoselov-2004}, the field of two-dimensional materials has grown 
immensely \cite{yazyev_mos2_2015,novoselov2016,chemrev-tan-2017,%
kang_2d_2020,mcp-joseph-2023}. One prominent group of such materials is given by
transition metal dichalcogenides (TMDCs) \cite{yazyev_mos2_2015}, consisting of
a layer of transition metal atoms (e.g., tungsten, molybdenum) sandwiched 
between two layers of chalcogenides (e.g., sulfur, selenium, tellurium). 
\ce{MoS2}, a member of this group, is a typically n-doped semiconductor with a 
strong spin-orbit interaction. In its monolayer form, broken inversion symmetry 
causes spin split bands \cite{yazyev_mos2_2015}. In addition, even in the 
monolayer limit it can be driven into intrinsic superconductivity via ionic 
doping \cite{lu2015,nnano-costanzo-2016}; for hole conduction, theory predicts 
it to be a topological superconductor 
\cite{hsu2017,prb-oiwa-2018,prb-wang-2022}.

Many attempts have been made to define quantum dots (QDs) in planar TMDC
materials \cite{jing_gate-controlled_2022}. However, the typically large
effective electron mass in the conduction band requires minuscule devices sizes
at the limits of traditional lithography, and most observations so far are
limited to classical, metallic Coulomb blockade \cite{li2013a,lau2022,song2015,%
song2015b,lee2016b} and QDs at defects \cite{papadopoulos2020,devidas2021a,%
krishnan2023a}. Only very recently quantization effects have been observed in
lithographically defined systems \cite{davari2020,kumar2023}. TMDC-based
nanotubes \cite{nature-tenne-1992} could naturally provide strong confinement 
in an additional dimension as well as perfect electronic boundary conditions 
compared to lithographically defined nanoribbons.

The challenging fabrication of long and defect free \ce{MoS2} nanotubes
\cite{remskar1996}, higher radii compared to CNTs \cite{seifert2002}, and the
fact that only multiwall nanotubes have been isolated up to now, have so far
limited research. Additionally, the TMDCs where stable and defect-free
nanotubes have been produced, are typically semiconductors. Metal contacts form
Schottky barriers, resulting in large contact resistances
\cite{zphys-schottky-1939}. Strong Fermi level pinning has been observed
\cite{nl-gong-2014,jpcc-sotthewes-2019}, further complicating the situation.
For planar TMDCs, recently remarkable advances in circumventing these barriers 
were made \cite{shen2021a,li2023,das2013}. Regarding nanotubes, for a long time
research was limited to optical and mechanical properties
\cite{serra_overview_2019,musfeldt_nanotubes_2020}, and work adressing
superconductivity in \ce{WS2} \cite{qin_diameter-dependent_2018}.
First attempts of low temperature transport spectroscopy used metals with a 
suitable low work function \cite{fathipour2015,reinhardt2019a}, but these 
metals were shown to react with and destroy the crystal lattice of \ce{MoS2}
\cite{reinhardt2019a,wu2019,remskar2022}. Only recently, using the semimetal
bismuth led to a breakthrough \cite{schock2023}.

Here, we utilize state of the art transfer techniques adapted from 2D materials
\cite{otsuka2021,castellanos-gomez2014} and classical semiconductor fabrication
to build electronic devices integrating \ce{MoS2} nanotubes. We compare these
techniques in terms of device yield and describe their effects on the device.
Overall, surprisingly, the devices fabricated with the classical ``Scotch
tape'' method alone perform better then any of the more sophisticated transfer
techniques. Additionally, we investigate several contact materials, extending
our previously published results \cite{schock2023}. Still bismuth remains so
far the best canditate for contacting \ce{MoS2} nanotubes.

\begin{figure}[tb]
\includegraphics{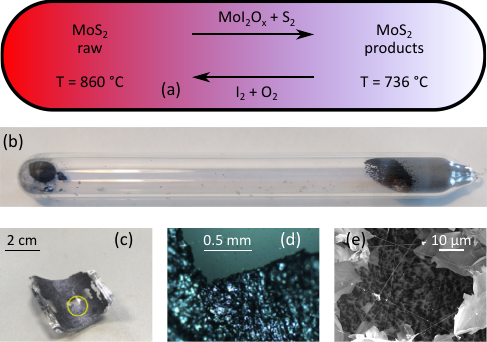}
\caption{
\ce{MoS2} nanotube growth: a) schematic of the chemical transport growth 
reaction, b) photograph of a growth ampoule containing the source material and 
the reaction result, c), d), optical images of the resulting material on a
piece of a broken ampoule, e) scanning electron micrograph of the material
displaying \MoSS\ nanotubes and flakes.
}
\label{fig:growth}
\end{figure}

\section{\ce{MoS2} nanotube growth}

In this work we use \ce{MoS2} nanotubes grown by an iodine-assisted chemical
transfer process \cite{remskar1996,nitsche_growth_1960,remikar_new_1998}. This 
technique utilizes the migration of the gas phase of a metal compound along a
temperature gradient from an area of vaporisation to an area of 
crystallisation. A halogen, in this case iodine, functions as transport agent.

The precursor, bulk crystalline \ce{MoS2}, is given into a quartz glass ampoule
together with iodine \ce{I2}, see Fig.~\ref{fig:growth}(a,b). The ampoule is
subsequently evacuated and sealed by locally melting the quartz glass. It is 
then heated up in a tube oven under presence of a temperature gradient, with 
the precursor \ce{MoS2} placed at the hotter end. Remaining oxygen \ce{O2} 
emitted from the quartz glass ampoule walls also participates in the reaction. 
The precursor 
reacts according to $\ce{MoS2} + \ce{I2} + \ce{O2} \longrightarrow \ce{MoI2O_x} 
+ \ce{S2}$, with the gaseous products then migrating along the temperature 
gradient to the cooler recrystallisation area, see Fig.~\ref{fig:growth}(a) 
\cite{nitsche_growth_1960}. There, the reverse process, $\ce{MoI2O_x} + \ce{S2} 
\longrightarrow \ce{MoS2} + \ce{I2} + \ce{O2}$, takes place. Subsequently the 
transport agents diffuse back to the hotter end, leading to a continuous 
process as long as the temperature gradient is maintained and feed material is 
present.

Over a growth period of approximately \SI{500}{\hour}, clean and long nanotubes
with a very low defect density form on the growth side, accompanied by
ribbon-like collapsed nanotubes, platelets, flakes, and more complex structures.
After a slow cool-down, the quartz glass ampoule is broken apart in order to 
access the grown and deposited material. Example images of growth results can 
be found in Fig.~\ref{fig:growth}(c,d,e).

\section{Transfer and assembly techniques}

Over the past decade, for research on 2D materials many different material
transfer and assembly techniques have been developed, and this process is still 
onging. In the following we discuss the usage of some of these methods for 
\ce{MoS2} nanotube devices. In general, more complex transfer methods are 
developed to avoid contamination by the transfer agents and achieve cleaner 
results. As an example, the anthracene crystal based method detailed below in
Section~\ref{sec:anthracene} has been used by Otsuka {\it et al.} with carbon
nanotubes, leading to photoluminescence spectra fully comparable to as-grown
macromolecules and thus indicating negligible contamination effects
\cite{otsuka2021}.

\begin{figure}[tb]
\includegraphics{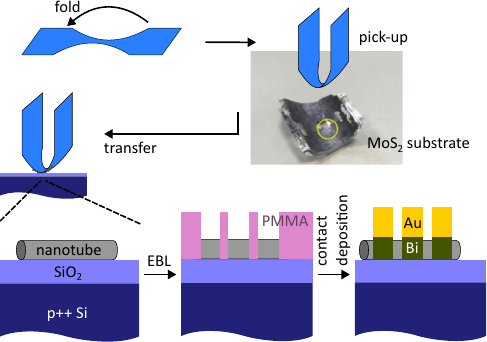}
\caption{
Schematic of the ``Scotch tape'' (or ``Nitto tape'') method as applied here.
First a piece of Nitto Denko ELP BT-150E-CM tape is cut into a strip narrowing
in the middle. Then the strip is folded and softly pressed onto the \ce{MoS2}
nanotube growth substrate, a piece of the original quartz ampoule. This
way nanotubes are picked up and can then be placed onto a receiving p-doped
Si-\ce{SiO2} substrate. Subsequently, using a standard electron beam
lithography (EBL) process, metal evaporation / sputtering, and lift off,
contacts are defined on the nanotubes.
}
\label{fig:bluetape}
\end{figure}

\subsection{The ``Scotch tape'' method}
\label{sec:scotch}

A very straightforward procedure to integrate \ce{MoS2} nanotubes into
electronic devices is the classical ``Scotch tape'', ``blue tape'', or ``Nitto
tape'' method initially developed for graphene \cite{science-novoselov-2004},
as illustrated in Fig.~\ref{fig:bluetape}. Nitto Denko ELP BT-150E-CM adhesive
tape is pressed onto the raw \ce{MoS2} material on a piece of the glass ampule.
Then the same piece of tape is pressed onto a Silicon wafer with a thermally
grown \SI{500}{\nano\meter} thick oxide layer and predefined chromium-gold
position markers. This randomly transfers nanotubes, flakes, and other \ce{MoS2}
nanostructures onto the chip surface at an adjustable surface density.
While the process is fairly straightforward and applies to a comparatively
large target area, also a small part of the adhesive coating is deposited,
leading to sticky patches and some contamination.

In order to contact specific nanotubes, their position on the chip is then
determined by optical microscopy. Surprisingly, even small diameter \ce{MoS2} 
nanotubes can be detected this way; we attribute the clear visibility to their 
outstanding optical properties \cite{remskar2022,kazanov_multiwall_2018}. The
optical images, including position markers, are used as base for the design of 
contact geometries. The chip is then spin-coated with poly-methyl metacrylate
(PMMA) resist; after a standard electron beam lithography process, the contact
metallization is deposited onto the exposed parts of the nanotubes and lift-off
in hot acetone is performed, see Fig.~\ref{fig:bluetape} and the discussion of
the different materials below.

\subsection{Suspending nanotubes between contacts}
\label{sec:suspension}

This process can be altered in order to reduce the disorder caused by the 
amorphous \ce{SiO2} surface and its surface charges; from carbon nanotubes it 
is well-known that suspending the nanostructures leads to significantly better 
spectroscopic results \cite{jarillo-herrero_electron-hole_2004,%
cao_electron_2005}. Prior to stamping the nanotubes onto sample, a PMMA resist
is deposited onto the surface. The stamp transfer works equally well for the
blank \ce{SiO2} surface and the hardened PMMA layer. After the transfer, a
second PMMA layer is then spincoated onto the first resist, such that the
nanotubes are ideally located at the interface between both layers. After EBL 
and development of the resist, the nanotubes are then suspended at the height 
of the interface between the two resist layers, held in place between
remaining, non-exposed resist areas. Following the the contact material
deposition and resist removal, they are then suspended between the contacts 
themselves.

The bottom resist layer was spin-coated using a ``\SIrange{1}{4}{\percent}
PMMA 50k'' solution, i.e., a solution of \SIrange{1}{4}{\percent} by weight of 
PMMA in anisol, where the average molecular weight of the PMMA polymer chains 
is 50000\,u. This resulted in a resist thickness range of
\SIrange{20}{60}{\nano\meter}, comparable to or less than the nanotube 
diameters. Alternatively, for a higher likelyhood of a finite gap between chip
substrate and nanotube, a \SI{9}{\percent} PMMA 200k solution, i.e., a solution
of \SI{9}{\percent} by weight PMMA of average molecular weight 200000\,u in
anisol, was used, leading to a resist layer thickness of about
\SI{200}{\nano\meter}. For the top layer in either case \SI{9}{\percent} PMMA
200k was applied. Different metallizations were tested, see also the discussion
and the SEM images in Fig.~\ref{fig:cracks}, Section~\ref{sec:sem} below, where
the quality of the resulting contacts is discussed. A typical thin contact
layer would consist of 40\,nm bismuth and 50\,nm gold, both thermally
evaporated; later experiments tested thicker but similar metallization layers.

\begin{figure}[tb]
\includegraphics{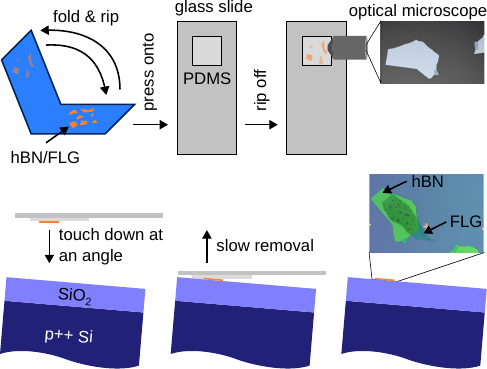}
\caption{
Schematics of the polydimethylsiloxane (PDMS) transfer method, mostly used for
few layer graphite (FLG) and hexagonal Boron nitrite (hBN) flake transfer.
First a piece of Nitto tape is cut into a strip and pressed upon the bulk
substrate of the transfer material, e.g., hBN. Then the strip is folded onto
itself and ripped apart repeatedly in order to thin down the material. By
pressing a piece of PDMS of $\sim \SI{1}{\cm} \times \SI{1}{\cm}$ onto the
Nitto tape and ripping it off, some of the flakes are transferred onto the
PDMS. After determining the position of a flake with an optical microscope, the
flakes are transferred onto a \ce{SiO2}/Si substrate.
}
\label{fig:pdmstransfer}
\end{figure}

\subsection{Polydimethylsiloxane (PDMS) transfer} Polydimethylsiloxane (PDMS) is 
a silicon based polymer widely used in 2D material science as a substrate and 
as a transfer agent \cite{castellanos-gomez2014,dong2024}. This is due to its 
flexibility and viscoelastic properties, which make the transfer of 2D flakes 
(and nanotubes) between different substrates possible.

Here, PDMS was primarily used to transfer quasi-two dimensional hexagonal Boron
nitrite (hBN) and few layer graphite onto \ce{SiO2} substrates with predefined
gold-structures on them, as highly conductive back gate and crystalline gate
isolator without dangling bonds. Literature on bilayer graphene has amply
demonstrated that this sort of material stack reduces disorder from the
amorphous surface of \ce{SiO2} as well as provides a very homogeneous electric
field \cite{engels2013,li2017a,nl-banszerus-2018,prx-eich-2018,overweg2018}.
While the transfer does not directly involve the \ce{MoS2} nanotubes, we
include it here for completeness.

Material transfer is achieved by at first pressing a ribbon of Nitto tape onto 
a bulk piece of the material in question, thereby retrieving a small amount of 
the 2D material, see Fig.~\ref{fig:pdmstransfer}. Subsequently, the tape is
folded, pressed together, and then ripped apart in order to break up the bulk 
stacks into smaller stacks and potentially monolayers of the 2D materials. This 
process is repeated at least 10 times. 

\begin{figure*}[tb]
\includegraphics{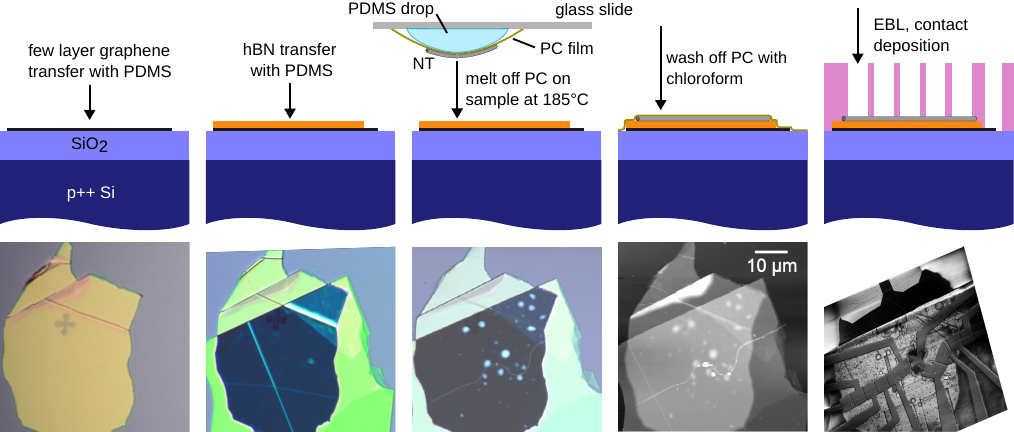}
\caption{
Schematics of the polycarbonate (PC) assisted transfer method. In the first
step, few layer graphene (FLG) and hBN are transferred with the ``PDMS method''
(see Fig.~\ref{fig:pdmstransfer}) and stacked upon each other. Then a
previously prepared strip of a thin PC film is placed and fixed upon a PDMS
droplet on a glass slide. In an xyz-stage, a nanotube is picked up from a
silicon wafer (prepared with the ``Scotch tape method'', see
Fig.~\ref{fig:bluetape}). The nanotube is placed upon the FLG-hBN heterostack,
and the PC is molten at \SI{185}{\degreeCelsius}. Afterwards, the PC is washed
off with chloroform and the nanotube is contacted through standard EBL
processing.
}
\label{fig:pdmspc}
\end{figure*}

A piece of a flat PDMS film, commercially available as Gelpak Gelfilm, is
attached to a glass slide and pressed onto the adhesive tape with the
exfoliated 2D materials. Using a very fast peel-off, some of the flakes are
transferred onto the PDMS film. The film is then inspected in an optical
microscope. Finally, flakes of the desired size and thickness are carefully
pressed onto a receiving \ce{SiO2} substrate and then remain on the surface
when peeling off the PDMS very slowly.

\subsection{Polycarbonate (PC) transfer}
\label{sec:pc}

This method, for targeted transfer of a nanostructure from one substrate to a
specific position on a different one, builds upon the two previous recipes.
A PDMS drop is used as stamp substrate, with a polycarbonate (PC) film covering
its surface to avoid direct contact between nanostructure and PDMS.

First, a thin layer of poly(bisphenol A carbonate) is fabricated through
coating a glass slide with a chloroform-PC solution of \SI{4}{\percent}. The
coating is subsequently air-dried. The resulting PC layer is slowly peeled off
the glass slide using an adhesive tape with a cutout area in the middle. The
PC-tape stack is then placed onto a pre-made PDMS droplet on another glass
slide, where it is secured such that the exposed PC layer is stretched over the
droplet. The resulting PC-film coated droplet is then carefully pressed onto a
substrate with an isolated, previously exfoliated nanotube or flake, heated to
130°C, and slowly peeled off, effectively picking up the material.

The resulting stack, consisting of the flake or nanotube, the PC layer, and the 
PDMS droplet on a glass carrier, is subsequently pressed onto a receiving
substrate and heated to 185°C, causing the PC to melt, see
Fig.~\ref{fig:pdmspc}. As the PDMS is lifted, the liquefied PC with the
attached nanomaterial structure remains on the sample. Finally, the PC layer is
gently washed off in hot chloroform for 15~minutes, leaving the nanotube or
flake on the substrate.

Compared to direct transfer with only PDMS, this procedure has two main
advantages. Due to the melting of the PC film, it is possible to transfer
materials which adhere stronger to PDMS than to the receiving substrate.
Further, as there is no direct contact between the flakes/nanotubes and the
PDMS, any contamination from PDMS is washed off together with the PC in the
chloroform rinsing step.

\begin{figure}[tb]
\includegraphics{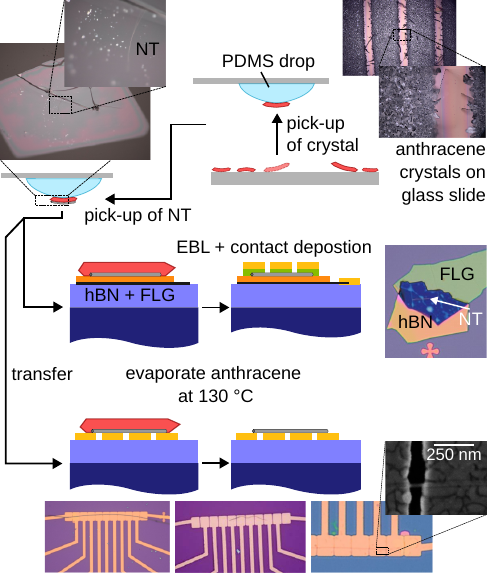}
\caption{
Anthracene crystal assisted transfer of a nanotube. First, anthracene crystals
are grown on a glass slide. Then, a large anthracene crystal is picked up with
a PDMS drop. With this, a nanotube, exfoliated with the scotch tape method, or
a 2D flake (hBN or FLG) can be picked up. These can then be either placed onto a
heterostack (middle part) or transferred onto predefined contacts (lower part).
The anthracene crystal with the material is pressed onto the chip and remains
there if peeled off slowly. After that, the anthracene is sublimated by heating
the device above \SI{130}{\degreeCelsius}.
}
\label{fig:anthracene-method}
\end{figure}

\subsection{Anthracene assisted transfer}
\label{sec:anthracene}

In case even the PC film transfer introduces too much contamination, a
recently developed replacement procedure utilizes anthracene crystals as
alternative
intermediate transfer agent between PDMS and the flakes or nanotubes
\cite{otsuka2021}. To implement this method, also illustrated in
Fig.~\ref{fig:anthracene-method}, we first grow anthracene crystals on glass
slides suspended \SI{1}{\mm} above granular anthracene heated to
\SI{80}{\degreeCelsius} in an ambient atmosphere. In order to grow large and
thin single crystals with a size of about \SI{1}{\milli\meter\squared},
following Otsuka {\it et al.} \cite{otsuka2021} a commercial permanent marker
was used to draw black lines on the glass slide, see
Fig.~\ref{fig:anthracene-method}. In the marked, dark regions, the growth of
crystals is suppressed; crystals growing nearby can extend above this region
and then reach larger sizes. The initial publication used a permanent marker of
type KOKUYO PM-41B; we found the type STAEDLER permanent Lumocolor S, Nr. 313-9
to be a suitable replacement. After a growth period of about \SI{12}{\hour},
large and homogeneous anthracene single crystals form on the slides.

With an optical microscope in a transfer setup, suitable crystals were chosen
and picked up with a PDMS droplet, see Fig.~\ref{fig:anthracene-method}.
Similar to the PC based method, the anthracene crystal was pressed upon a flake
or nanotube and the substrate was subsequently heated to a temperature of above
\SI{80}{\degreeCelsius} in order to increase the adhesion of the anthracene to
the object to be transferred. Then, the nanotube, anthracene, and PDMS stack
was rapidly peeled off (in under a second) to ensure that the anthracene
adheres more strongly to the PDMS than to the \ce{SiO2} chip surface. In order 
to deposit the nanomaterial at the top of the stack, it was pressed upon the
desired location of the receiving chip, heated to a temperature above
\SI{90}{\degreeCelsius}, and very slowly peeled off over a duration of about
one minute, which left the anthracene with the transferred stack on the surface
of the chip. The anthracene crystal was then sublimated by heating to a
temperature above \SI{130}{\degreeCelsius}; it typically leaves no visible
contamination residues.

As shown in Fig.~\ref{fig:anthracene-method}, this method allows for two 
different approaches to device fabrication. Either a heterostack can be
assembled with subsequent deposition of top contacts, or the nanotube can be
placed upon predefined contacts with trenches between them. As the anthracene
evaporates in an ambient atmosphere, without the need of any wet chemical 
processing, there is no danger of ripping off thin layers or nanotubes due to
surface tension \cite{otsuka2021}. Using predefined contacts is obviously
limited to contact materials which do not form an insulating oxide
barrier in ambient atmosphere, unless additional encapsulation steps are
performed.

\section{Contact engineering}

For all the deposition of nanotubes and/or layer assembly of devices, achieving
good electrical contacts to the nanomaterial is of central importance. In
particular, here we talk about transparent and non-destructive contacts
\cite{schock2023}: transparent meaning having a low resistance and Ohmic
behaviour, and non-destructive meaning that the contact fabrication does not
significantly damage the molecular and thereby electronic structure of the
nanomaterial.

For planar, quasi-twodimensional \ce{MoS2}, a large amount research has been
invested into this topic world-wide, with the primary objective of Ohmic
contacts for \ce{MoS2}-based field effect transistors. As with many other TMDC
materials, strong Schottky barriers typically form at the semiconductor-metal
interface \cite{zphys-schottky-1939,lince_schottky-barrier_1987}. The precise
mechanisms involved in their formation have long been under discussion. While
the mismatch of the metal work function certainly contributes, see the
discussion below, additionally strong Fermi level pinning takes place at the
interface \cite{monch_valence-band_1998,nl-gong-2014,kim_fermi_2017,%
jpcc-sotthewes-2019}.

In the case of \ce{MoS2} nanotubes, the reduced geometry poses additional
difficulties. An edge contact to a \ce{MoS2} flake is effectively a
one-dimensional interface; the same edge contact to a nanotube however
zero-dimensional. Further, while graphene has been shown to make good contacts
to planar \ce{MoS2} and other TMDC \cite{chanana2016}, so far no such success
has been achieved by depositing a ``flat'' graphene or graphite layer onto a
``round'' nanotube or vice versa --- an observation which can likely be
attributed to the shape mismatch.

\subsection{Impact of the contact material}

Several different approaches to avoid the formation of a Schottky barrier at
metallic contacts to \ce{MoS2} have been proposed so far. Primarily these
revolve around the selection of the contact material. In classical
semiconductor technology, a Schottky barrier is minimized by adapting the metal
work function to the semiconductor. For \ce{MoS2} with an electron affinity of
$\lambda_{\ce{MoS2}} = 4.0\,\text{eV}$, this means selecting a {\em low-work
function metal} such as titanium or scandium \cite{das2013,reinhardt2019a,%
wu2019}. In the following we name this a {\em type-I contact}. As
demonstrated in \cite{reinhardt2019a}, scandium can be used to contact
\ce{MoS2} nano\-tubes, however, serial charge traps make Coulomb blockade
spectroscopy difficult. In hindsight, the origin of these charge traps is
obvious -- the chemically reactive metal destroys the \ce{MoS2} layer structure
already during deposition \cite{wu2019}.

Inserting a thin insulating layer (e.g., an insulating hBN monolayer) as a
{\em transparent tunneling barrier} which at the same time prevents Schottky
barrier formation has been attempted with some success on 2D materials
\cite{chen2013,dankert2014,farmanbar2016a,pande2020}. Accordingly this was also
tested for \ce{MoS2} nanotubes, see the discussion below, and named {\em type-II
contact}. Local doping in the contact areas is another technique transferred
from existing semiconductor technology \cite{gao2020,jiang2023}. For nanotubes,
the small relevant surface area makes this difficult to implement; in addition,
surface dopants immediately lead to potential irregularities that would pose
problems in low temperature measurements. Since copper doping has shown promise
in other works \cite{liu_improvement_2018}, we also have tested bulk copper
contacts.

Recently, it was discovered, that the use of {\em semi\-metals} was a promising
way to avoid the formation of strong Schottky barriers at a \ce{MoS2} interface
\cite{shen2021a}. In the interface region of a semiconductor and a metal,
hybridization of the electronic bands of the semiconductor and the metal leads
to so-called metal-induced-gap-states (MIGS) and via them to Fermi level pinning
\cite{louie_electronic_1976}. The density of states (DOS) of a semimetal
however approaches zero at the Fermi level, leading to a corresponding
reduction in MIGS. This in turn reduces stability of the Schottky barrier and
makes Ohmic contacts possible \cite{shen2021a,li2023,schock2023}. Graphene has
already been used to contact planar \ce{MoS2} \cite{chanana2016}, attempts with
\ce{MoS2} nanotubes have failed so far however, most likely due to the mismatch
in shape / geometry. This leads us to the elements bismuth
\cite{shen2021a,schock2023} and antimony \cite{li2023,lee2024}, already highly
successful for planar materials, as contact layers. This approach is in the
following named {\em type-III contact}.

\begin{figure}[tb]
\includegraphics{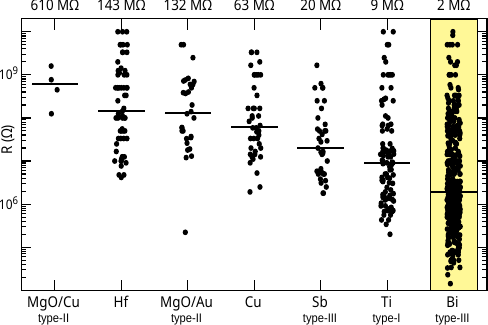}
\caption{
Two terminal resistances of \ce{MoS2} nanotubes and nanoribbons contacted with
different contact materials. Each point in the graph corresponds to the room
temperature resistance measured between two contacts on a nanotube or
nanoribbon. Additionally, the median resistance for each material is marked
with a black line. Clearly, bismuth leads to the smallest median resistance, an
order of magnitude smaller than the second best tested material titanium. Part
of the data has already been shown in \cite{schock2023}.
}
\label{fig:resistances-materials}
\end{figure}

The devices prepared in this work were measured under ambient conditions. The
two-point resistance was determined by applying a constant bias voltage of
\SI{10}{\milli\volt} and measuring the resulting current. The resulting values 
for different contact materials are shown as scatter plots in
Fig.~\ref{fig:resistances-materials}, with the median of each material given as
horizontal line and written out above the corresponding column. We do not
differentiate between deposition methods here, though the large majority of
devices was fabricated using the simplest ``Scotch tape'' transfer,
either on the chip surface (see Sec.~\ref{sec:scotch}) or on a pre-deposited
resist layer (see Sec.~\ref{sec:suspension}). Bismuth-based semi-metal contacts
(i.e., {\em type-III}) clearly outperformed all other tested materials, with a
median two-point resistance value of $R_\text{Bi} = 2\,\text{M}\Omega$
\cite{shen2021a,schock2023}. This confirms corresponding work on planar
\ce{MoS2} as well as our own previous publication \cite{schock2023}.

The material with the second best median two-point resistance was titanium.
It is a metal with a well suited work function for n-type conduction band
contacts to \ce{MoS2} (i.e., {\em type-I}), however has also shown to be highly
reactive \cite{wu2019}. Similar results and disadvantages have been seen for
scandium \cite{reinhardt2019a}. As such, even though the room temperature
results are promising, charge traps and potential irregularities at the contacts
will likely make low-temperature transport spectroscopy challenging.

Aside bismuth, also the semimetal antimony has been used to successfully make
contacts to planar TMDC materials \cite{li2023}. However, for our nanotubes the
results using antimony were already significantly worse, with a median value
$R_\text{Sb} = 20\,\text{M}\Omega$. Note that Li {\it et al.} only achieved
their best contact resistances with antimony grown in the ($01\bar{1}2$)
direction, depositing the contacts under elevated temperatures of about
\SI{100}{\degreeCelsius} \cite{li2023}. In contrast, here, the antimony was
deposited at room temperature, and no clear statement on its crystal
orientation was possible. In addition, compared to planar \ce{MoS2}, where a
topmost layer can be matched by the crystal structure of the contact material,
nanotubes expose a curved surface to the semimetal. This likely prevents
even locally the formation of a single crystalline layer with a matching
structure.

Further material combinations tested include gold and copper -- as
well-conducting metals with a large electronic density of states -- on top of a
thin MgO tunnel barrier, i.e., {\em type-II} contacts. As visible in
Fig.~\ref{fig:resistances-materials}, in both cases the resulting two-point
resistances are comparatively high. Even given the scatter and the still
relatively small number of data points, further investigations seem not
worthwhile, with median two-point resistance values of $610\,\text{M}\Omega$ and
$132\,\text{M}\Omega$.

\begin{figure}[tb]
\includegraphics{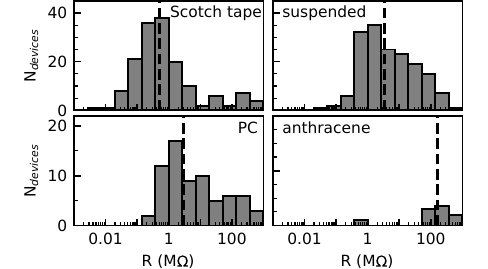}
\caption{
Two terminal resistance distributions of bismuth-contacted nanotube devices,
for four different material transfer methods -- ``Scotch tape'' deposition
(Section \ref{sec:scotch}), suspended nanomaterial (Section
\ref{sec:suspension}), polycarbonate-assisted deposition (Section
\ref{sec:pc}), and anthracene-assisted deposition (Section
\ref{sec:anthracene}).
}
\label{fig:resistances-techniques}
\end{figure}

\subsection{Impact of the transfer technique}

In Fig.~\ref{fig:resistances-techniques}, we compare the observed two terminal
resistances of devices prepared with different transfer techniques. In all
cases, bismuth has been used as the contact layer. As clearly visible from the
figure, so far, the simplest ``Scotch tape'' transfer technique shows the best
results, with a much larger number of devices exhibiting two terminal
resistances below \SI{1}{\mega\ohm}. Both suspending the nanomaterial (see
Sec.~\ref{sec:suspension}) and PC transfer (see Sec.~\ref{sec:pc}) seem to have 
a negative impact on the fraction of devices produced with a two-point 
resistance below \SI{1}{\mega\ohm}. Additionally, much more devices with 
resistances above \SI{10}{\mega\ohm} can be observed, indicating a reduced 
quality of the contacts.

The validity of the evaluation is limited insofar as the fabrication using
more complex methods took mostly place at a later time and used later
nanomaterial growth batches; a hypothetical change in clean room chemicals
quality, raw material properties (which we do not have any further indications
of), etc., would materialize similarly in the plots.

As for the anthracene method, only a very small number of devices was tested so 
far. All the data in Fig.~\ref{fig:resistances-techniques} for anthracene
devices stems from contacts to a total number of 4 nanotubes prepared in two
different ways; it can only tentatively indicate that this approach also
reduces contact quality. The chips with device resistances above
\SI{100}{\mega\ohm} were prepared with contacts predefined by EBL procedure and
a channel length of about \SI{100}{\nano\meter}. In the two devices on one
nanotube with resistances below \SI{1}{\mega\ohm}, the nanotube was transferred
onto gold contacts pre-sliced with the beam of a focused ion beam (FIB) system.
The channel length in these devices was at about \SI{50}{\nano\meter} much
smaller than that of other device types, which may contribute to slightly lower
device resistance. Further, it is conceivable that the Ga\textsuperscript{+}
ions deposited by the FIB beam in the vicinity of the trenches dope the surface
of the nanotubes and therefore lower the Schottky barrier. With only one
nanotube tested so far, these ideas are however only speculative and require
further investigation.

\begin{figure} 
  \includegraphics{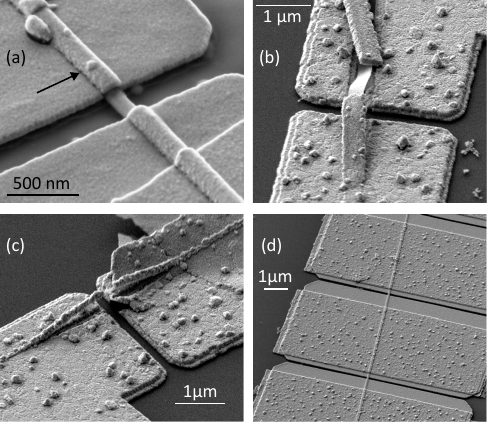}
  \caption{SEM images of devices after fabrication and probe station
    characterization. (a) Nanotube contacted with a bismuth layer of
    \SI{40}{\nm} and a gold layer of \SI{50}{\nm}; (b) nanoribbon and (c)
    nanotube on the same sample contacted with a bismuth layer of \SI{50}{\nm}
    and a gold layer of \SI{100}{\nm} both evaporated under two different
    angles, resulting in a step-like perimeter; (d) nanotube transferred with
    anthracene onto \SI{100}{\nm} gold contacts with $\approx \SI{100}{\nm}$
    gaps and subsequently contacted with \SI{25}{\nm} bismuth and \SI{30}{\nm}
    gold, both evaporated at two different angles.
  }
  \label{fig:cracks}
\end{figure}

\subsection{SEM failure analysis}
\label{sec:sem}

The large scatter of the resistance data points in
Fig.~\ref{fig:resistances-materials} and Fig.~\ref{fig:resistances-techniques}
even in the case of nominally identical device preparation indicates that the
fabrication process still has fundamental limitations. To identify problems,
several devices were imaged in detail in a scanning electron microscope; 
example results can be seen in Fig.~\ref{fig:cracks}.

A possible cause of the reduced contact quality is the occurrence of gaps or
cracks between the contact material deposited on top of the nanotube and the
contact material deposited next to them, see, e.g., the black arrow in
Fig.~\ref{fig:cracks}(a). While not occuring for carbon nanotubes due to their
much smaller radius, this is a well-known (and sometimes intended) phenomenon
when evaporating materials onto elevated structures such as semiconductor
nanowires. Due to the directional material deposition via thermal evaporation,
a contact material 'cap' forms on the top of the nanotube. As the cap grows, it
begins to shadow a larger area below and next to the nanotube, reducing
deposition in that region. A precise adjustment of the contact material
thickness to the nanomaterial dimensions, in order to counteract this, is in our
case challenging since typically nanotubes and ribbons of varying size are
deposited on the same chip.

Initial resistance testing of devices suggested an optimum of the electronic
behaviour around the contact layer thickness used to fabricated the device of
Fig.~\ref{fig:cracks}(a) \cite{schock2023}. However, as explained above, if the
contact material films are not thick enough, the cap of contact material on top
of the nanostructure and the surrounding contact material may not consistently
reach each other. As found out subsequently, this is possible in the region 
indicated with an arrow in the SEM image of Fig.~\ref{fig:cracks}(a), 
indicating the need for a more robust contact metallization.

In addition, the SEM images of devices suggest grain-based growth of the 
contact films, especially for bismuth, with corresponding fluctuation of the 
layer thicknesses and potential gap sizes. Even in presence of a gap, contact 
could still occur occasionally. This is not limited to the suspended nanotube 
case, but is very likely to be more do\-mi\-nant there, which could further 
explain the wide scatter of contact resistances for all materials and the 
slightly worse results for devices intended to feature suspended nanotubes.

In order to prevent the gap formation, the contacts deposited in
Fig.~\ref{fig:cracks}(b) were significantly thicker; in addition, thermal
evaporation of the contact materials was done at two separate angles of device
orientation to improve coverage. Preliminary results indicate that this could
at least partially improve the contact quality; one of two devices fabricated
so far had several contacts with resistances below \SI{1}{\mega\ohm}.
The second one, with the nanoribbon depicted in Fig.~\ref{fig:cracks}(b),
performed much worse, clearly since almost all nanoribbons and -tubes were
ripped out of the embedding material. The reason for this is unclear so far.

\section{Discussion of the resistance scatter}

Of all tested contact materials, the bismuth-gold combination so far remains
the most promising one -- in combination with the least sophisticated
fabrication method. Nevertheless, the observed two-point resistances scatter
widely. What is the cause?

A {\em wide distribution of nanomaterial properties} cannot be fully excluded.
The \MoSS\ growth process delivers flakes, nanotubes, nanoribbons as well as
breathing and twisted variants \cite{remskar1996,eliseyev2023}. During the
transfer process, long, straight, and thin structures are preferably selected
in the optical microscope. Distinguishing nanotubes from nanoribbons and
determining the precise dimensions would however require time-consuming SEM or 
AFM imaging, which may also lead to additional contamination or damage.

{\em Contamination during lithography}, as, e.g., organic resist residues or
reactions with photoresists or -developers, is another relevant topic. Moving
the fabrication away from optical lithography to electron beam lithography with
its organic chemistry only has so far not led to clear improvements. Imperfect
dissolution of resist layers during development is possible. While the
resist remainders can in principle be removed with a brief oxygen plasma based
descum process, the plasma treatment will strongly attack the sulphur surface
and have an impact on its own
\cite{bolshakov_contact_2019,mahlouji_contact_2021,lee_effects_2017}.

{\em PDMS or PC contamination during stamping} can reduce the device quality. 
It is already known that PDMS contaminates the surface of 2D materials after the
stamping process \cite{otsuka2021,schwartz2019,jain2018a}. As both FLG and hBN
were transferred using PDMS this can introduce disorder in the heterostack,
degrading the quality of the backgate. Additionally, contamination on the hBN
surface is in direct contact with the nanotubes and could influence its
electronic properties. Finally, during the PC transfer of the nanotube, 
residues of PDMS may spread onto the contact suface of the nanotube. Together 
with contamination from the PC itself \cite{schwartz2019}, these would be 
directly at the nanotube-semimetal interface.

{\em Insufficient metallization or metallization gaps} at the contacts can, as 
discussed already above, particularly affect large-diameter nanomaterials and
suspended structures. One may also speculate that too thick bismuth regions 
become non-conductive at low temperature, leading to additional resistive 
barriers. Metal deposition under varying angles and thicker gold cap layers 
should be used to mitigate these effects.

{\em Tear-off of the nanomaterial} has been observed, e.g., in
Fig.~\ref{fig:cracks}(b,c), as also discussed above; while the nanotube or 
nanoribbon remains whole, it entirely or in part lifts out of the contact 
electrodes, taking part of the material with it. Again thicker metallization, 
here combined with a more careful lift-off procedure, may be required. Surface 
tension during drying would pull a possibly suspended nanomaterial towards the 
substrate; we expect this to lead to different types of damage. Nevertheless, 
also the use of a critical point dryer may be considered for future devices.

\section{Conclusions and outlook}

\MoSS\ nanotubes and nanoribbons have significant potential for quantum 
electronic devices. Here, we compare different contact materials and material 
transfer techniques and their effect on the contacts to these \ce{MoS2} 
nanomaterials. 

Regarding the contact materials, we can so far conclude that the best choice 
for \ce{MoS2} nanotubes and nanoribbons is bismuth \cite{shen2021a,schock2023}, 
a semi-metal leading to the minimzation of metal-induced gap states and Fermi 
level pinning. While, for planar \ce{MoS2}, antimony, also a semimetal, has led
to record conductivities \cite{li2023,lee2024}, this could not be confirmed for 
nanotubes. The comparison of different transfer techniques indicates that the 
classical ``Scotch tape'' method \cite{science-novoselov-2004} in its simplicity 
still gives the most reliable results. We tentatively conclude that more 
complex fabrication procedures still pose more danger of surface contamination. 
A large scatter of measured resistance values remains, which can be due to 
several different causes. Insufficient coverage of the non-planar nanomaterial 
and the formation of minuscule ``nano-gaps`` between the contact material 
covering the nanotubes and the contact material surrounding them, even at the 
apparent optimal layer thickness, seems to play an important role, with surface 
contaminations secondary in effect. 

In order to reduce the impact of the nano-gaps, multiple-angle evaporation 
as well as an overall thicker layer of contact material was used. First data 
indicate an improved likelihood of good contacts. Evaporation onto a heated 
device substrate as well as annealing are further approaches to be followed in 
the future. Regarding the reduction of potential surface contamination, \ce{O2} 
and \ce{Ar} plasma treatments shall be tested as next steps
\cite{bolshakov_contact_2019,mahlouji_contact_2021,lee_effects_2017}, as
well as \ce{H2S} exposure of the devices at elevated temperature 
\cite{peto_spontaneous_2018}. Even though clearly not all approaches apply to
nanotubes and nanowires, the highly active world-wide research on planar
\MoSS\ field effect transistors provides a multitude of avenues to follow.

\begin{acknowledgement}
We gratefully acknowledge funding by the DFG via grants Hu 1808/4-1 (project id
438638106) and Hu 1808/6-1 (project id 438640730) and by the Slovenian Research
Agency via grant P1-0099. We would like to thank Lain-Jong Li for insightful 
discussions and Ch.~Strunk and D.~Weiss for the use of experimental facilities. 
The measurement data was recorded using Lab::Measurement \cite{labmeasurement}.
\end{acknowledgement}

\bibliographystyle{pss}
\bibliography{paper}

\newpage

\section*{Graphical Table of Contents\\}
GTOC image:
\begin{figure}[h]%
\includegraphics[width=4cm,height=4cm]{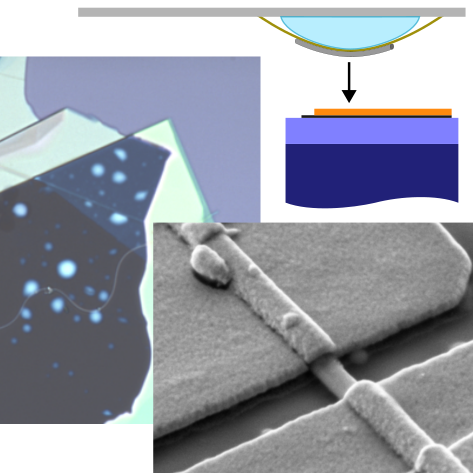}
\caption*{%
Clean and defect-free MoS\textsubscript{2} nanotubes are highly promising as
novel material for quantum electronic devices. With this in mind, we look at
device fabrication methods originally developed for planar, 2D transition
metal dichalcogenides, and discuss their implementation for single
MoS\textsubscript{2} nanotubes and -ribbons. This includes various transfer and
assembly methods as well as the choice of contact materials.
}
\label{GTOC}
\end{figure}

\end{document}